\documentclass[aps,pre,showpacs,floatfix]{revtex4}
\usepackage{amsmath,bm,epsfig}
\usepackage{latexsym}
\usepackage{amssymb}
\usepackage{color}
\usepackage{morefloats}
\usepackage{subfigure}
\usepackage{graphicx}
\usepackage{amsmath}
\usepackage{bm}
\usepackage{epsf,epsfig,graphics}
\usepackage{float}
\usepackage{makeidx}

\begin{document}

\title{Polymers in fluid flows}
\author{Roberto Benzi$^1$ and  Emily S.C. Ching$^2$}
\affiliation{$^1$Department of Physics, Univ. Roma "Tor Vergata";
email: benzi@roma2.infn.it} \affiliation{ $^2$Department of
Physics, The Chinese University of Hong Kong, Shatin, Hong Kong;
email: ching@phy.cuhk.edu.hk}

\date{\today}
\begin{abstract}
The interaction of flexible polymers with fluid flows leads to a
number of intriguing phenomena observed in laboratory experiments,
namely drag reduction, elastic turbulence and heat transport
modification in natural convection, and is one of the most
challenging subjects in soft matter physics. In this paper we
review our present knowledge on the subject. Our present knowledge
is mostly based on direct numerical simulations performed in the
last twenty years, which have successfully explained, at least
qualitatively, most of the experimental results. Our goal is to
disentangle as much as possible the basic mechanisms acting in the
system in order to capture the basic features underlying different
theoretical approaches and explanations.

\vspace{0.2cm} \noindent Keywords: polymer-flow interaction, drag
reduction, elastic turbulence, heat transport modification

\end{abstract}

\maketitle

\vspace{0.5cm}

\noindent Posted with permission from the Annual Review of
Condensed Matter Physics, Volume 9 \copyright  2018 by Annual
Reviews, http://www.annualreviews.org/.

\section{Introduction}\label{intro}
The interaction of flexible polymers with fluid flows is one of
the most challenging subjects in soft matter physics. Laboratory
experiments show that even a minute concentration of polymers can
dramatically change properties (drag reduction) of turbulent flows
or, if the flow is laminar, can trigger a new form of turbulence
named elastic turbulence. How to properly describe flow-polymer
interactions and how to understand the basic physical phenomena
observed in experiments is both a fundamental scientific issue and
an important challenge to develop  many industrial applications.
In this paper we review our present knowledge on the subject. In
the last twenty years, direct numerical simulations of
polymer-flow dynamics have successfully explained, at least
qualitatively, most of the experimental results. Starting from
this consideration, our goal is to disentangle as much as possible
the basic mechanism acting in the system. Many different theories
have been proposed and we feel confident that future work can
provide a reasonably unified approach on the subject.

\section{POLYMER FLOW INTERACTION}

\subsection{Polymer passive advection}

 We start our
review by understanding the dynamics of polymers advected by a
turbulent  fluid flow.  A flexible polymer can be considered as a
chain of $N \gg 1$ monomers.  The size of a polymer in its
stretched configuration is of the order of tens of micrometers. In
most cases, this size is much smaller or at most equal to the
smallest dissipative scale of turbulence, namely the Kolmogorov
scale $\eta = (\nu^3/\epsilon)^{1/4}$, where $\nu$ is the
kinematic viscosity of the fluid and $\epsilon$ is the rate of
turbulent energy dissipation. In this section we consider a
polymer to be a passive object advected by the flow velocity and
neglect its feedback to the flow. The important information comes
from the end-to-end vector $\bf R$ of the polymer configuration.
In its simplest form the Lagrangian dynamics of $\bf R$ is
described by~\cite{Birds1987,lebedev1,lebedev2}
\begin{equation}
\label{1.1}
\partial_t { R_i}  = - \frac{1}{2\tau} { f(R) R_i} + { R_j}  { \partial_j} { u_i} + \sqrt{ \frac{R_0^2}{ \tau } } w_i(t)
\end{equation}
where $u_i$ is the component of the velocity field, $\tau$ is the
polymer relaxation time, $R_0^2$ is a parameter that takes into
account the role of thermal fluctuations, and $w_i(t)$ is
independent white noise, delta-correlated in time. The relaxation
time $\tau$ depends on the chemical and physical properties of the
polymer, and ranges from $10^{-3}$~s up to tens of seconds. The
function $f(R)$ takes into account the finite extensibility of the
polymer and that the maximum extension of polymer size $R=|{\bf
R}|$ is $R_{\rm max}$. Hereafter we choose $f(R)=1/(1-\alpha
R^2)$, with $\alpha = 1/R_{\rm max}^2$. Using Equation~\ref{1.1},
we can describe the statistical properties of polymers by
considering the conformational tensor $R_{ij} = \langle R_i R_j
\rangle_{w}$ where the average is done on the noise. Upon using
Ito calculus, the Eulerian equations for the conformational tensor
are:
\begin{equation}
\label{1.2} \frac{d R_{ij}}{dt} \equiv \partial_t R_{ij} + u_k
\partial_k R_{ij} = - \frac{1}{\tau} \left[ f(R) R_{ij}-
\delta_{ij} R_0^2 \right] + R_{ik}\partial_k u_j +
R_{jk}\partial_k u_j
\end{equation}
First we consider the case $\alpha=0$, assuming that $R < R_{\rm
max}$ at any time.  If  we  follow the polymer along its
trajectory, the quantities $s_{ij} \equiv \partial_j u_i$ are a
function of $t$ only. Starting from Equation~\ref{1.1}, we are
interested in computing the statistical properties of $R$.
Following References\cite{lebedev2,chertkov,vincenzi1}, we can
compute the probability distribution $P(R)$ from the knowledge of
the long time behavior of the Lyapunov exponent of the Lagrangian
trajectory. Let ${\bf l}(t)$ be the solution of the linear
equation:
\begin{equation}
\label{1.3} \frac{dl_i}{dt} = s_{ij} l_j
\end{equation}
with initial condition $ |{\bf l}(0)| = 1$. For chaotic flows, the
size $l(t) = |{\bf l}(t)|$ grows exponentially in time as
$\exp({\gamma t})$ where $\gamma $ is the finite time Lyapunov
exponent, i,e, $\lim_{t \rightarrow \infty} \gamma = \lambda$,
$\lambda$ being the Lyapunov exponent of the flow. It is also
known \cite{benzi1} that the probability $P(\gamma)$ of observing
the value  $\gamma$ at time $t$ is given by
$\exp[-tS(\gamma-\lambda)]$, where $S(x)$ is known as the Cramer
entropy and is a convex function with a minimum at $x=0$. The
Cramer entropy can be used to estimate the large deviation
properties of $l(t)$. More precisely, one can show that $\langle
l^q(t) \rangle = \exp[tL(q)]$ where $L(q)$ is the Legendre
transform of $S(x)$, i.e. $L(q) = \sup_{x}[qx-S(x)]$.
Asymptotically, we can estimate from Equation~\ref{1.1} that $R
\sim \exp(-t/\tau+\gamma t)$ and  that  stretching properties of
the polymer depend on the dimensionless number Wi$_\lambda ~\equiv
\lambda \tau$. The quantity Wi$_\lambda$ is called Weissenberg
number, although in many papers no reference to the Lyapunov
exponent is made. Clearly there exists a critical value ${\rm
Wi}_{\lambda,c}$ of ${\rm Wi}_\lambda$ for which the polymer tends
to stretch indefinitely (for the case of $\alpha=0$). This
critical value signals the so-called coil-stretch transition for
the polymer.  It is possible to show \cite{lebedev2,chertkov}
that, for ${\rm Wi}_\lambda < {\rm Wi}_{\lambda,c}$, the polymer
extension $R$ shows a power law distribution $P(R) \sim R^{-1-a}$
where $a = 2 \tau L(a)$. The quantity $L(q)$ is not known in
general and depends on the intermittent features (if any) of the
Lagrangian trajectory. In most cases one can assume as a first
approximation $L(q) = \lambda q + \Delta q^2/2$ and obtain
\begin{equation}
\label{1.4} a = \frac{\lambda}{\Delta} \left[ \frac{1}{{\rm
Wi}_\lambda} -2 \right]
\end{equation}
From Equation~\ref{1.4} we can immediately see that $ {\rm
Wi}_{\lambda,c} = 1/2$: For $ {\rm Wi}_\lambda < {\rm
Wi}_{\lambda,c}$ the value of $a$ is positive and the probability
distribution $P(R)$ is normalizable; for ${\rm Wi}_\lambda > {\rm
Wi}_{\lambda,c}$ the exponent $a$ becomes negative and $P(R)$ is
no longer normalizable. The ratio $\lambda/\Delta$ is a
quantitative measure of intermittency for the Lagrangian
trajectories: Strong intermittency implies $\lambda/\Delta \ll 1$
and small values of $a$;  small intermittency is equivalent to
$\lambda/\Delta \gg 1$ and large values of $a$.

For a large value of Wi$_\lambda$, we  must consider the case
$\alpha
>0$ in Equation~\ref{1.1}. In this case, the analysis, performed
in \cite{vincenzi2} and \cite{vincenzi3} in some simplified case,
shows that coil-stretch transition does not change qualitatively,
although the existence of a scaling range for $P(R)$ depends on
flows details. The overall message coming from the above
discussion is rather clear: A single polymer passively advected in
a chaotic flow undergoes a coil-stretch transition when
Wi$_\lambda$ is larger than a critical value ${\rm
Wi}_{\lambda,c}$. The precise value of ${\rm Wi}_{\lambda,c}$
depends on the flow properties [i.e. the function $L(q)$],
although it is reasonable to guess it is of order one. So far one
has not been able to measure $\lambda$ experimentally. For fully
developed turbulence, one may use the simple estimate $\lambda
\sim 1/\tau_{\eta}$ where $\tau_{\eta}= \sqrt{\nu/\epsilon}$ is
the Kolmogorov time. Alternatively, one can estimate $\lambda$ as
$U/L$ where $U$ is a characteristic velocity of the flow and $L$
its characteristic scale. Quite often the Weissenberg number
defined by ${\rm Wi}=U\tau/L$ is also referred to as the Deborah
number, ${\rm De}$. Not surprisingly, depending on the definition,
the critical value of ${\rm Wi}$ or ${\rm De}$ changes
considerably. However, the basic point of the above analysis is
that the time criterion based on ${\rm Wi}_\lambda
> {\rm Wi}_{\lambda,c}$ is the one to be used to capture the polymer coil-stretch
transition.

\begin{figure}[h]
\includegraphics[width=4in]{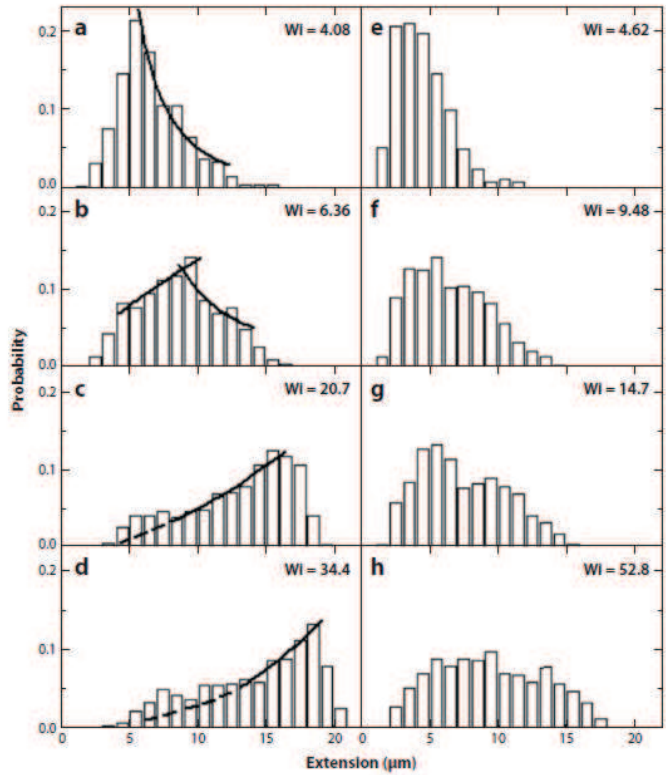}
\caption{Probability distribution $P(R)$ of the end-to-end polymer
length $R=|{\bf R}|$ measured in a swirling flow. Black lines are
best fit with power-law distribution $R^{-1-a}$: A clear
coil-stretch transition is observed for ${\rm Wi} = {\rm Wi}_c
\sim 6$. For ${\rm Wi} < {\rm Wi}_c$, $P(R)$ shows a power-law
distribution with $a \sim 1.5$ for the smallest ${\rm Wi}$ in
panel (a). Adapted from Reference~\cite{steinberg1} with
permission.} \label{fig1}
\end{figure}

$P(R)$ and coil-stretch transition have been investigated
numerically \cite{boffetta1,Gotoh2010,Schumaker2006} and
experimentally in Reference \cite{steinberg1} (see
Figure~\ref{fig1}). In the experiment, polymers were advected in a
swirling flow between the edge of a uniformly rotating glass of
radius $r_1$ and angular velocity $\Omega$. The Weissenberg number
is defined as ${\rm Wi}=\tau \Omega r_1/d$ where $d$ is the gap
between plates. The authors considered polymers with $R_0 \sim 1
\mu$m and $R_{\rm max} \sim 20 \mu$m and they were able to measure
the value $R$ by fluorescent techniques. A clear coil-stretch
transition is observed for ${\rm Wi} = {\rm Wi}_c \sim 6$. For
${\rm Wi} < {\rm Wi}_c$, $P(R)$ shows a power-law distribution,
and $a \sim 1.5$ for the smallest Wi studied. The experiment shows
a rather remarkable qualitative agreement with the theoretical
framework discussed in this section.

\subsection{When turbulence is affected by polymers}
After discussing how polymers are affected by a fluid flow, we now
discuss the change in the fluid flow due to the presence of
polymers. We consider three-dimensional homogeneous and isotropic
turbulent flows far away from boundaries and review the ideas of
Lumley~\cite{Lumley73} and de Gennes\cite{TabordeGennes,deGennes}
on when turbulence would be affected by polymers.

For three-dimensional homogeneous and isotropic turbulence, the
Kolmogorov 1941 theory holds approximately with corrections due to
intermittency. Energy is injected at the large scales, and
cascaded down to smaller scales, and eventually dissipated at the
Kolmogorov scale $\eta$ at which molecular viscosity directly
acts. At each scale $r$ in the inertial range, there is a
characteristic fluctuating velocity $\delta u(r)$, related to $r$
by
\begin{equation} \frac{[\delta u(r)]^3}{r} \approx \epsilon  \label{2.1} \end{equation}
A crude estimate of the stretching at scale $r$ can be obtained by
$\delta u(r)/r \equiv 1/\tau_r$. Upon using Equation \ref{2.1}, we
obtain $\tau_r \sim r^{2/3}\epsilon^{-1/3}$. Our discussions in
Section~2.1 lead one to think that polymers would affect
turbulence at scales $r$, where $\tau/\tau_r \ge 1$. Thus for
scales $r \le r^*$, where $r^*=(\epsilon \tau^3)^{1/2}$ is defined
by $\tau_{r^*}=\tau$, turbulence would be affected in a certain
way. This is the idea originally proposed by Lumley. According to
this idea, polymers would affect the turbulent flow as long as
$r^* \ge \eta$, regardless of the concentration of the polymers.

De Gennes followed a completely different approach. He thought
that polymers could produce effect on turbulence only when the
elastic energy stored by the polymers becomes comparable to the
turbulent kinetic energy. He assumed that at scale $r$, the
elastic energy of polymers per unit volume would decrease with $r$
as
\begin{equation} E_{el}(r) =  c
k_B T \left(\frac{r^*}{r}\right)^{5m/2} \label{elastic}
\end{equation}
where $c$ is the concentration of polymers per unit volume, $k_B$
is the Boltzmann constant, $T$ is the temperature of the polymer
solution, and $1 \le m \le 2$ is some positive exponent depending
on the specifics of the flow. Note that in Equation \ref{1.1}, the
elastic energy is taken to be quadratic in $R$, i.e. $m=4/5$. The
turbulent energy per unit volume at scale $r$ is given by $\rho
[\delta u(r)]^2$, where $\rho$ is the density of the polymer
solution and increases with $r$. Thus polymers would affect
turbulence at scales $r\le r^{**}$, where the scale $r^{**}$ is
the scale at which the elastic energy and the turbulent kinetic
energy become equal. Using Equations~\ref{2.1} and \ref{elastic},
and the expression for $r^*$, we obtain
\begin{equation}
r^{**} = \left(\frac{c k_B T}{\rho} \right)^v \epsilon^{1/2-v}
\tau^{3/2-v} , \label{r**} \end{equation} where $v =
(5m/2+2/3)^{-1}$. For very small $c$, $r^{**}$ is smaller than
$\eta$, and polymers cannot affect turbulence. Thus there exists a
threshold concentration for the polymers to have any effect on
turbulence. For dilute polymer concentrations above the threshold
value, $\eta < r^{**} < r^*$ and as concentration increases
$r^{**}$ approaches $r^*$. De Gennes argued that polymers should
be considered as passive for $r^{**} < r < r^*$ and the turbulent
energy cascade would stop at $r^{**}$, so $r^{**}$ can be thought
of as an effective cutoff scale replacing the Kolmogorov scale
$\eta$. However, he provided no information on how the turbulent
energy is eventually dissipated.

Thus the theory of de Gennes allows us to see clearly how and why
the concentration of polymers must play a role in their effects on
turbulence, whereas Lumley's idea leads to the scale $r^*$, which
is a possible upper-bound scale for polymers to affect turbulence.
However, neither of the two theories gives any details on how
turbulence is affected.  Several experiments have been carried out
to study the change in statistics of a turbulent counter-rotating
disk flow by
polymers~\cite{NJP2008,Ouellette2009,Xi2013,Quitry2016}.
Turbulence statistics such as Eulerian velocity structure
functions are modified in the presence of polymers. It is found
that the Eulerian velocity structure functions can be made to
collapse into one master curve in Reference~\cite{Xi2013} but into
two families, one for low polymer concentration and another for
higher concentration in Reference~\cite{Quitry2016}, when the
separation is rescaled by some length scale that depends on
polymer concentration $c$ and relaxation time $\tau$. The precise
form of this length scale is
 different in these experiments. Suggestions to relate this length scale to
$r^{**}$ or a new length scale at which the turbulent kinetic
energy flux and the elastic energy flux are equal have been made
\cite{Xi2013}. However, we have yet to obtain a full understanding
of all these experimental results.

\section{POLYMER EFFECT ON FLOWS: EXPERIMENTS}
\subsection{Drag Reduction}

One well-known effect is polymer reduces friction drag in
turbulent wall-bounded flows. This effect was discovered by
Toms~\cite{Toms}, who observed that an addition of about 10 parts
per million by weight of a long chain polymer (polymethyl
methacrylate) can lead to a significant reduction of friction drag
in a turbulent flow of monochlorobenzene in a pipe while studying
the degradation of polymers. Similar effect has been observed in
turbulent pipe or channel flow of water with polyethylene oxide
 or polyacrylamide. A large number of experimental
studies have been carried out to characterize this phenomenon (see
e.g. Reference \cite{Virk75} for a review). Below, we will
summarize the key features of the phenomenon.

For a fluid of density $\rho$ flowing in a pipe of diameter $D$,
the friction drag is measured by the dimensionless Fanning
friction factor $f_D$, defined as the ratio between wall shear
stress $\tau_w$, which is the work done per unit volume due to the
applied pressure gradient, and the kinetic energy density of the
mean flow
\begin{equation}
f_D \equiv \frac{\tau_w}{\rho U_{av}^2/2} = \frac{D \Delta
p}{2\rho U_{av}^2 L} \ . \label{3.1}
\end{equation}
Here $\Delta p$ is the pressure drop across a distance $L$ in the
pipe, and $U_{av}$ is the mean velocity, averaged over time and
the cross-section of the pipe, and is given by $\int_0^{D/2} V(y)
2 \pi (D/2-y) dy/(\pi D^2/4)$, where $V(y)$ is the streamwise
velocity averaged over time and $y$ is the distance from the wall.
 When polymers are added, $f_D$ is not changed at low Reynolds number ${\rm Re} = U_{av} D/\nu$.
At a certain value Re$_c$, the onset of drag reduction occurs and
$f_D$ is reduced. Experimental measurements show that the onset
value Re$_c$ depends on polymer concentration~\cite{Nadolink1987}.
The reduction of $f_D$ is tantamount to an increase in the mean
streamwise velocity $V(y)$ for a given pressure drop $\Delta p$ or
a decrease in the pressure drop required for a fixed mean
streamwise velocity. Such an enhancement of $V(y)$ by polymers is
shown in Figure~\ref{fig2} in terms of the commonly used wall
units, defined by $V^+(y)=V(y)/u_\tau$ and $y^+=y u_\tau/\nu$,
where $u_\tau=\sqrt{\tau_w/\rho}$ is the friction velocity. After
the onset, $f_D$ decreases when the polymer concentration is
increased, and $V^+(y)$ approaches an apparently universal
limiting curve that cannot be exceeded by increasing the
concentration further. This limit is referred to as the maximum
drag reduction (MDR) asymptote (see Figure~\ref{fig2}a). Theories
of drag reduction will be discussed in section 5.2.

\subsection{Elastic Turbulence}

Another interesting effect is that polymer with sufficiently long
relaxation time or high Weissenberg number Wi can give rise to
 an irregular flow state with
velocity fluctuations spanning a broad range of spatial and
temporal scales even at low Reynolds number. This irregular flow
state at high Wi and low Re is known as elastic
turbulence~\cite{Steinberg2000} and is caused by an instability
due to the polymer stresses. The first experiments of elastic
turbulence were performed in dilute polymer solution in flows with
curved streamlines such as swirling flow between two plates,
Taylor-Couette flow, and flow in a curvilinear
channel~\cite{Steinberg2004}. More recently, elastic instability
has also been observed experimentally in a long, straight
microchannel~\cite{Pan2013}.

\begin{figure}[h]
\includegraphics[width=5in]{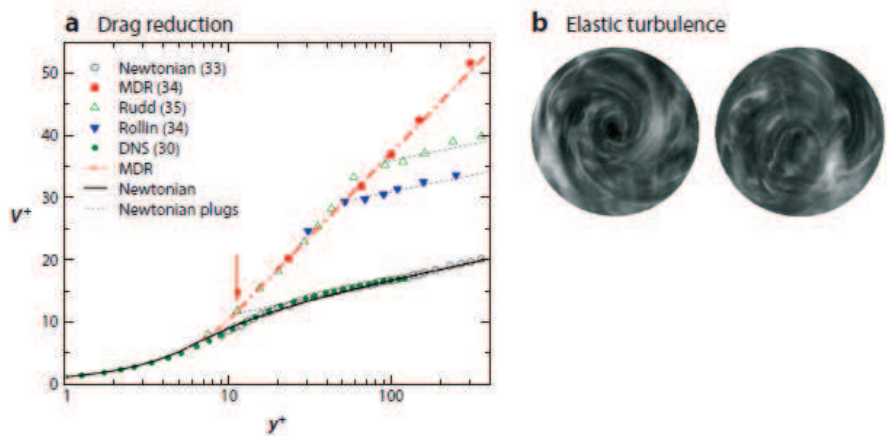}
\caption{(a) A synthetic view of experimental and numerical
results on drag reduction in wall bounded turbulence. The
horizontal axis represents in log scale the value of  $y^+$,
whereas the vertical axis shows the mean streamwise velocity $V^+$
in wall units. The black solid line represents the Newtonian
profile. The red dot-dashed line represents the MDR asymptote (see
also~\cite{Virk75}) and is a representation of Equation
\ref{5.25}. By increasing the polymer concentration for fixed
pressure drop, there is a systematic increase of drag reduction up
to the asymptotic MDR regime. Reprodiced from
Reference~\cite{Procaccia2008} with permission. (b) Elastic
turbulence. Two representative snapshots of the polymer flow at
Wi=13 and Re=0.7 taken from below. The field of view corresponds
to the upper plate area. The flow was visualized by seeding the
fluid with light reflecting flakes. Reproduced from
Reference~\cite{Steinberg2000} with permission. Abbreviations:
DNS, direct numerical simulations; MDR, maximum drag reduction.}
\label{fig2}
\end{figure}

For example, in a swirling flow between two plates, the fluid in a
stationary cylindrical cup of flat bottom plate is set into
swirling motion by a co-axial rotating upper plate just touching
the surface of the fluid. The average shear stress $\sigma$ is
measured as a function of the average shear rate $\dot \gamma$.
For small Weissenberg number, defined by ${\rm Wi}=\tau \dot
\gamma$, $\sigma$ remains close to the value $\sigma_{\rm
lam}=\eta(\dot \gamma) \dot \gamma$ for a laminar shear flow at
the same angular velocity $\omega$, where $\eta(\dot \gamma)$ is
the viscosity of the polymer solution.  It is found that when Wi
is above some critical value, $\sigma/\sigma_{\rm lam}$ increases
significantly and there is a transition from the laminar flow to
elastic turbulence. For the same range of shear flow rates of the
pure solvent without polymers, the ratio $\sigma/\sigma_{\rm lam}$
remains at unity within resolution of the measurements. The
frequency power spectrum of the velocity fluctuations in elastic
turbulence displays a power-law decay, which spans over about a
decade in frequencies. The power-law dependence indicates a broad
range of time scales of the motion and resembles that of developed
turbulence of Newtonian fluid at high Re but the energy spectrum
displays a steeper slope indicating that the flow is random but
not as ``rough" as in the case of fully developed turbulence of
Newtonian fluid. The fluctuating velocity field can be visualized
in the snapshots of the flow of the polymer solution above the
transition as shown in Figure~\ref{fig2}b.

\section{EQUATION OF MOTION}
\begin{figure}[h]
\includegraphics[width=4in]{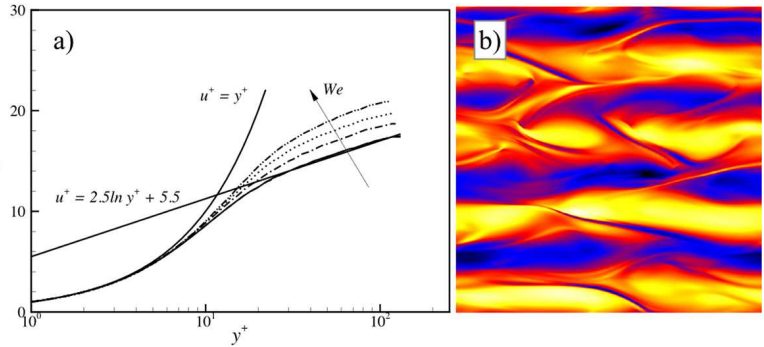}
\caption{(a) Velocity profile of the streamwise velocity $V^+$ in
Prandtl units, obtained in DNS of the Oldroyd-B model for
different values of Weissenberg number, (here denoted by  We);
${\rm We} = \tau S_0$, where $S_0$ is the value of the shear at
the boundary $y^+=0$. By increasing $\tau$ or We, one observes a
clear decrease in the overall drag. Reproduced from
Reference~\cite{Taegee2003} with permission. (b) Snapshot of the
vorticity field obtained in
 a two-dimensional simulation  of
Kolmogorov flow with polymer at Re=0.48 and Wi=31. The system
shows a clear instability leading to a turbulent-like behavior
similar to the phenomenon of elastic turbulence. Adapted from
 Reference~\cite{Boffetta2008new} with permission.} \label{fig3}
\end{figure}

Given Equation~\ref{1.2} for the conformational tensor $R_{ij}$,
the effect of polymers on the flow can be computed using the
momentum balance, which reads as
\begin{eqnarray}
\label{4.1}
\partial_t u_i +  u_j \partial_j u_i &=& - \frac{1}{\rho} \partial_i p +  \partial_j T_{ij} + \nu_s \Delta u_i  \\
\label{4.2}
T_{ij}  &\equiv& \frac{ \nu_p}{\tau} \left[ f(R)
\frac{ R_{ij} }{ R_0^2 }-  \delta_{ij} \right]
\end{eqnarray}
Here, $\nu_s$ is the kinematic viscosity of the solvent and
$\overline{\overline{\mathbf{T}}}$ (with components $T_{ij}$)
represents the momentum stress tensor due to polymer, where
$\nu_p$ is the polymer contribution to the zero-shear viscosity
$\nu_0=\nu_s+\nu_p$ of the polymer solution and $\nu_p$ increases
with polymer concentration $c$~\cite{Birds1987}. Equation
\ref{4.1} is often presented by writing $\nu_s = (1 - \beta)
\nu_0$ and $ \nu_p = \beta \nu_0$, which is convenient for
numerical studies with a constant $\nu_0$. The overall free energy
$F$ in the system is defined as
\begin{equation}
\label{4.3} F = \int d {\bf r} \left\{  \frac{\rho u^2}{2} +
\frac{\nu_p}{\tau} [ {\rm tr} R_{ij} - R_0^2 \ln \, {\rm det}
(\overline{ \overline{ \mathbf{R}}}/R_0^2) ] \right\}
\end{equation}
Using Equations \ref{1.2} and \ref{4.2}, one can obtain the energy
dissipation,
\begin{equation}
\label{4.4} \frac{dF}{dt} = \int d {\bf r} \left[\{
\frac{\nu_s}{2} (\nabla u)^2 + \frac{1}{\tau} {\rm tr} [ \overline
{\overline{\mathbf{T}}} (1+\overline{
\overline{\mathbf{T}}})^{-1}\overline{ \overline{\mathbf{T}}}]
\right\}
\end{equation}
Equations  \ref{1.2} and \ref{4.1} with \ref{4.2} are known as the
finite extensible nonlinear elastic model with Peterlin's
approximation, or the FENE-P model in short, for dilute polymer
solutions. The FENE-P model is based on the assumption of a simple
nonlinear elastic force describing polymer stretching whereas more
complex physical effects are not taken into account (see, for
instance Reference \cite{Celani2006}) and  it is always
challenging to compare the relaxation time in the FENE-P model
against real experiments. Nevertheless the FENE-P model can
qualitatively capture the relevant experimental observations
described in the previous section, namely elastic turbulence and
drag reduction. If one ignores the finite extensibility of the
polymer, $f(R)=1$ and the resulting equations are usually referred
to as Oldroyd-B model. In the past few years,  there have been a
number of direct numerical simulations (DNSs) of the  FENE-P or
Oldroyd-B model for wall bounded turbulence
\cite{Taegee2003,Sureshkumar1997,Dimitropoulos1998,DeAngelis2004,Ptasinski2003}
(see also \cite{White2008} for a detailed review). In Figure
\ref{fig3}a, we show the numerical results obtained in
\cite{Taegee2003}  for the stream wise velocity profile for
different values of Weissenberg number (here denoted by We). There
is a clear increase in the velocity or a decrease in drag as We
increases. The numerical results should be compared against the
laboratory measurements shown in Figure \ref{fig2}a. This is clear
evidence that the FENE-P or Oldroyd-B model captures the
qualitative behavior of drag reduction found in laboratory,
although a quantitative agreement may be available only after fine
tuning of the model parameters. Two-dimensional numerical
simulations have been studied in References
\cite{Boffetta2003,Boffetta2008new} for a Kolmogorov flow with
periodic boundary conditions. The system is forced with a periodic
forcing and attains a velocity field $ (U_0 \cos(2 \pi y/L), 0)$
where $L$ is the size of the system. It is known that for ${\rm
Re} \equiv U_0 L/\nu_0$ smaller than the critical value ${\rm
Re}_c = \sqrt{2}$, the flow is stable when there are no polymers.
By increasing the Weissenberg number ${\rm Wi}= \tau U_0 / L$, the
system develops an instability and the dynamics becomes chaotic in
time and space, i.e. the system shows the characteristic features
of elastic turbulence. Figure \ref{fig3}b, as well as a more
detail analysis given in Reference \cite{Boffetta2008new}, clearly
shows that elastic turbulence can be explained (at least
qualitatively) by the constitutive Equations (\ref{1.2}) and
(\ref{4.1}).

There is a simple but highly non-trivial argument that allows us
to understand the role of the concentration parameter $\nu_p$ in
the FENE-P model~\cite{Benzi2004, Procaccia2008}. We consider the
case where there is  substantial stretching such that
$\delta_{ij}$ in Equation \ref{4.2} can be neglected. Let us
assume that the concentration changes as $\nu_p \rightarrow n
\nu_p$ where $n$ is any positive real number. This is equivalent
to saying that $R_{ij} \rightarrow R_{ij}/n$ and, more
importantly, the function $f(R)$ changes as
\begin{equation}
\label{4.5} f(R) \rightarrow \frac{1} {1 - (\alpha/n^2) R^2} \ ,
\end{equation}
which shows that by changing the concentration we (formally)
change the value of the maximum polymer extension $R_{\rm max}$ in
the FENE-P equations. Next, we observe that it is possible to
define an effective polymer relaxation time
\begin{equation}
\tau_{\rm eff} \equiv  \frac{\tau}{f(R)} \ , \label{tau_eff}
\end{equation} which gives
\begin{equation}
\label{4.6} \tau_{\rm eff}(n) = \tau  \left( 1- \frac{\alpha}{n^2}
R^2 \right)
\end{equation}
Equation \ref{4.6} tells us two pieces of information: (a) For $ n
\gg 1$ we obtain $\tau_{\rm eff} = \tau$, i.e. for high
concentration the FENE-P model becomes independent of the
concentration itself; (b) for $ n \ll 1$ even a very small amount
of stretching leads to $\tau_{\rm eff} \to 0$. In Section 2.1, we
see that polymer stretching occurs for ${\rm Wi}_\lambda = \lambda
\tau \sim 1$, thus we can safely assume that the effect of
polymers on the flow becomes relevant when ${\rm Wi}_{\rm eff}
\equiv \tau_{\rm eff}(n) \lambda$ is large enough. This is
equivalent to say that there exists a critical value $n_c$ or a
critical concentration, for which $\tau_{\rm eff}(n_c) \lambda
\sim 1$: For $n<n_c$ or concentration below the critical value,
the effect of polymers is negligible, whereas for $n
> n_c$ polymers can change the (turbulent) properties of the flow.
In other words, the FENE-P model seems to capture the basic
argument due to de Gennes on the existence of a critical
concentration for polymer effects on flows, without assuming any
critical scale and/or any cascade argument for the energy flux. As
a side result, for the Oldroyd-B model with $f(R)=1$, the effect
of the concentration disappears and any change in the
concentration is equivalent to a redefinition of $R_{ij}$ with no
effect on the dynamics; the only relevant parameter being the
polymer relaxation time $\tau$. This implies that changing the
polymer relaxation time $\tau$ in Oldroyd-B model is qualitatively
equivalent to changing concentration parameter $\nu_p$ for fixed
$\tau$ in the FENE-P model. The above discussion is independent of
the way turbulence is generated and, in particular, of any wall
effect.

\section{DRAG REDUCTION}

\subsection{Drag reduction without drag}

\begin{figure}[h]
\includegraphics[width=5in]{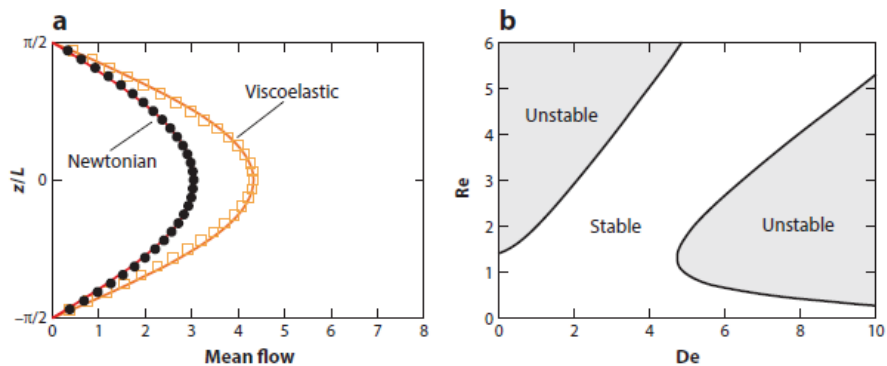}
\caption{(a) Comparison between the Newtonian profile and the
Oldroyd-B simulations for a Kolmogorov flow with periodic forcing
and mean flow $U \cos( z/L)$ with ${\rm De} = U\tau/L \sim 4$. A
clear increase in the mean flow is observed for the polymeric
case. Adapted from Reference \cite{Boffetta2005a} with permission.
(b) Theoretical computation of the stability region in the plane
Re-De for the Kolmogorov flows. Two different regions are
observed: At small De the critical Reynolds number ${\rm Re}_c$
increases for increasing De, i.e. the effect of polymers
stabilizes the onset of hydrodynamical instability; at large De,
${\rm Re}_c$ decreases, i.e. instabilities are driven by
polymer-flow interaction (elastic turbulence). Adapted from
Reference \cite{Boffetta2005b} with permission.} \label{fig4}
\end{figure}

The basic feature of fully developed three-dimensional turbulence
is the Re-independence of the energy dissipation rate $\epsilon$.
This feature, sometimes referred to as the zeroth law of
turbulence, is often written as
\begin{equation}
\label{5.1} \epsilon = C_D \frac{U^3}{L}
\end{equation}
where $U$ represents the  typical turbulent velocity and $L$ its
corresponding scale. Notwithstanding the  arbitrary definition of
$U$ and $L$, the constant $C_D$ has the physical meaning of the
 drag coefficient. To see that this is the case, we can
compute $\epsilon$ as $FU/\rho$ where $F$ is the forcing per unit
volume acting on the system. From Equation \ref{5.1} we obtain:
\begin{equation}
\label{5.2} C_D = \frac{FL}{\rho U^2}
\end{equation}
which is precisely the definition of the friction factor in
wall-bounded turbulence~(see Equation~\ref{3.1}). Using Equation
\ref{5.1} and/or Equation \ref{5.2}, one can study the phenomenon
of drag reduction for systems with no boundaries. Here we show
some of the many cases studied in recent years. We start by
reviewing the results obtained for three dimensional turbulent
flow for a Kolmogorov forcing with periodic boundary conditions
and mean flow $U\cos(z/L)$. We considered the limit of large
Re=$UL/\nu$,  and relatively small ${\rm De}=U \tau/L$. In Figure
\ref{fig4}a, we show the main result obtained in
\cite{Boffetta2005a}  for ${\rm Re} \sim 200$ and ${\rm De} \sim
4$. The comparison between the Newtonian case (${\rm De}=0$) and
the effect of polymers (${\rm De} \sim 4$) clearly shows that
there is an increase of the mean flow due to polymers.  A direct
computation of $C_D$ given by Equation \ref{5.2} shows a
significant amount of drag reduction by almost a factor of two.
Similar results have been obtained for turbulent flows forced by
constant shear~\cite{Brasseur2010}. Following Reference
\cite{Boffetta2005b}, the case of Kolmogorov forcing can be
studied analytically for small Re and De. The stability diagram
for the Oldroyd-B model is displayed in Figure \ref{fig4}b. Two
clear regions are present: For small Re and  De, the critical
value ${\rm Re}_c$  of the Reynolds number increases with
increasing De, i.e., the flow is more stable due to the effect of
polymers; for small Re and large De a second region of unstable
flows appears and corresponds to elastic turbulence discussed in
Section 3.2. Using multiple-scale analysis, it is possible to
compute analytically the drag coefficient $C_D$ for relatively
small Re and De~(see Reference \cite{Boffetta2005c}). One obtains
$C_D({\rm De}) = C_D(0)/(1+b {\rm De})$ with $b$ a positive
constant, i.e. drag tends to decrease as De is increased. Although
the above analytical results do not refer to turbulent flows, they
clearly show that the phenomenon of drag reduction should be
considered in more general terms and not only restricted to
wall-bounded turbulent flows. The computation of the drag
coefficient can also be done for homogeneous and isotropic
turbulence. In Reference \cite{Deangelis2005}, DNSs of the FENE-P
model have been performed for different values of De. In
Table~\ref{table1}, we show the value of $C_D$ (rightmost column)
as a function of De (first column): Clearly for large enough De,
we observe the effect of drag reduction (in the way defined in
this section).  To perform a scale by scale analysis for
homogeneous and isotropic turbulence with polymers
\cite{Deangelis2005},  we consider $R_{ij} = B_i B_j$, where $B_i$
is a vector field \cite{lebedev2}. The equation for $B_i$ and
$u_i$ are given by
\begin{eqnarray}
\label{5.3}
\partial_t u_i + u_j \partial_j u_i &=& - \frac{\partial_i p}{\rho} + \frac{\nu_p}{\tau}\partial_j [f(B)B_i B_j] + \nu_s \Delta u_i ,\\
\label{5.4}
\partial_t B_i + u_j \partial_j B_i &=& -\frac{1}{\tau}[f(B)B_i - 1]
+B_j \partial_ju_i ,
\end{eqnarray}
where $f(B) = 1/(1-\alpha B_i B_i)$ with the repeated index
summation notation used. The relative advantage in using Equations
\ref{5.3} and \ref{5.4} is that we can easily compute the analog
of the $4/5$ equation in this case, namely:
\begin{equation}
\label{5.5} \langle \delta u(r)^3 \rangle + \frac{\nu_p
f(B)}{\tau} \langle \delta u(r) \delta B(r)^2 \rangle = -
\frac{4}{5} \epsilon r - \frac{\nu_p f(B)}{\tau^2} \int_r dr'
\langle \delta B(r')^2 \rangle,
\end{equation}
where $\delta u(r) = ({\bf u}( {\bf r+x },t) - {\bf u} ({\bf
x},t)) \cdot {\bf r}/r$, $\delta B(r) = ({\bf B}( {\bf r+x},t) -
{\bf B} ({\bf x},t)) \cdot {\bf r}/r$ and $r = | {\bf r}|$. We
identify the de Gennes length scale $r^{**}$ as $l_p$, the value
of $r$ at which the two terms on the left-hand side of Equation
\ref{5.5} are equal. Similarly we can define the scale $r_e$ as
the value of $r$ at which the two terms on the right-hand side are
equal. If $r_e \sim l_p$, we obtain
\begin{equation}
\label{5.7} \frac{\delta u(l_p) }{l_p} \sim \frac{1}{\tau_{\rm
eff}},
\end{equation}
where we have used Equation~\ref{2.1}. In other words the de
Gennes scale $l_p$ satisfies the generalised Lumley criteria given
by Equation \ref{5.7}. Using DNS \cite{Deangelis2005} it is
possible to show that $r_e \sim l_p$, as reported in
Table~\ref{table1}. The overall picture is that, within the FENE-P
model, there exists a unique scale $l_p$ satisfying both the de
Gennes criterion and the generalized Lumley criterion Equation
\ref{5.7}. Hereafter, we shall refer to the scale $l_p$ as the
Lumley-de Gennes scale. As discussed in the previous section,
$l_p$ depends on the concentration. We remark that something
similar to the scale $l_p$ has been extracted from experimental
data in Reference \cite{Xi2013} as discussed in Section 2.2.

\begin{table}[h]
\tabcolsep7.5pt \caption{Parameters of the numerical simulations
discussed in \cite{Deangelis2005}} \label{table1}
\begin{center}
\begin{tabular}{@{}l|c|c|c|c@{}}
\hline
 De & $\epsilon_T$ & $r_e$ &  $l_p$ & $C_D = \epsilon_T/(u_{rms}^3/L)$ \\
\hline
0      & 0.156 & - & -  & 1.87 \\
0.18 & 0.174 & 1.23 & 0.603 & 1.90 \\
0.54 & 0.238 & 2.20 & 0.973 & 1.55 \\
0.54 & 0.232 & 2.20 & 0.923 & 1.60 \\
\hline
\end{tabular}
\end{center}
\end{table}

\begin{figure}[h]
\includegraphics[width=5.5in]{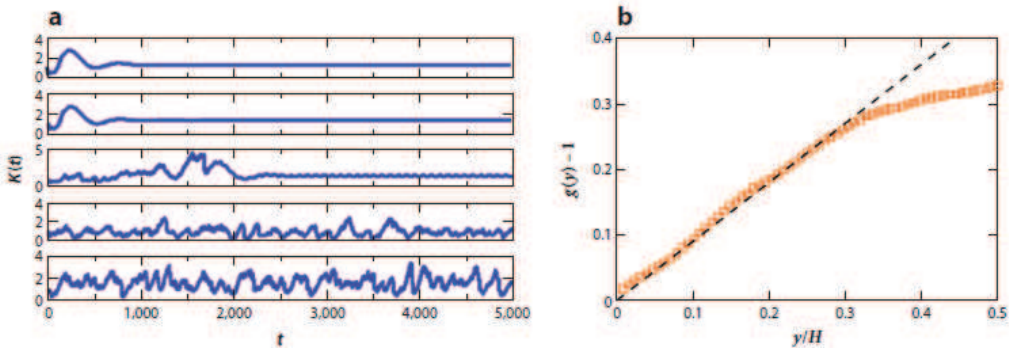}
\caption{(a) Elastic turbulence in shell models. The Weissenberg
number Wi is increasing from the top to the bottom: At low Wi the
model is not chaotic, whereas by increasing Wi the model exhibits
chaotic behavior due to polymer-flow interaction. Adapted from
Reference~\cite{Ray2016} with permission. (b) Computation of the
effective viscosity $g(y^+)$ in wall-bounded turbulence. The
effective viscosity is computed as the ratio of energy dissipation
rate due to polymer and energy dissipation rate due to turbulent
fluctuations: $g(y^+)-1$ exhibits a linear profile (dashed line)
close to the wall. Panel b courtesy by A. Scagliarini.}
\label{fig5}
\end{figure}

Equations \ref{5.3} and \ref{5.4} can be used to define a shell
model \cite{Benzi2004,Benzi2003,Pandit2005,Benzi2004b}. It is
worth noting that shell models have been extensively used to
successfully understand  some features of the statistical
properties of intermittency in three dimensional turbulence. In
this case, shell models are able to reproduce the relevant
statistical features of DNS of homogenous and isotropic turbulence
for FENE-P model, including drag reduction as previously defined.
Also shell models can be used to demonstrate the existence of
elastic turbulence at extremely low Re, as recently shown in
\cite{Ray2016}~(see Figure~\ref{fig5}a. In short, shells model
can be used as a simple numerical tool to understand the basic
features of energy fluxes end energy exchange in the FENE-P model
and provide a clear cut definition for the Lumley-de Gennes scale
$l_p$.

\bigskip

\subsection{Drag reduction in wall bounded flows}

We now turn our attention on the phenomenon of drag reduction in
wall-bounded turbulence. Our major interest is to understand how
the average velocity profile changes due to polymer effects in
turbulent flows, as described in section 2.1. Our starting point
is to briefly review the von Karman theory for wall-bounded
turbulence. Using wall units defined in section 3.1,  we can
define $H^+ \equiv Hu_{\tau}/\nu = {\rm Re}_{\tau} \gg 1$ ,  $S^+
= \partial_y V(y) \nu /u_{\tau}^2$ the dimensionless shear of the
average flow $V^+$ and $W^+ = - \langle w u \rangle/u_{\tau}^2$
the (absolute) turbulent momentum flux. Then momentum conservation
reads:
 \begin{equation}
 \label{5.20}
 S^+ + W^+ = 1- \frac{y^+}{{\rm Re}_{\tau}}
 \end{equation}
Hereafter we shall neglect the last term on the right-hand side of
Equation \ref{5.20}. To close the problem, we need a relation
between $S^+$ and $W^+$ in agreement to the von Karman hypothesis.
It turns out \cite{Procaccia2008} that such a phenomenological
equation can be derived using the energy balance in the form of
\begin{equation}
\label{5.21} \left[ \frac{\delta^+}{y^+} \right]^2+ \frac{
\sqrt{W^+} }{\kappa_v y^+} = S^+
\end{equation}
Upon multiplying Equation \ref{5.21} by the turbulent kinetic
energy $K^+ \sim W^+$ , we can interpret as follows:
  the first term on left-hand side is the turbulent dissipation near the viscous layer $y^+= \delta^+ \equiv \nu/u_{\tau}$;
  the second term on the left-hand side is the turbulent energy dissipation obtained by von Karman original argument while
  the term on the right-hand side is the turbulent energy production. Solution of Equations  \ref{5.20} and \ref{5.21}
  gives the Newtonian profile (black line) shown in Figure \ref{fig2} panel (b) with $\delta^+=5.7$ and $\kappa_v = 0.4$.
  Asymptotically, we have $V(y^+) = (u_{\tau}/\kappa_v)\log(y^+) +A$ where $A \sim 5.5$. The region between the asymptotic behavior
  of $V(y^+)$ and $\delta^+$ is usually referred to as the buffer layer, i.e., a region near the wall where the maximum of turbulent energy production is located.

Lumley \cite{Lumley73} and de Gennes \cite{deGennes}  provided
different theoretical interpretations for drag reduction in
wall-bounded turbulent flows ( see also \cite{White2008} for a
recent review). Lumley attributed drag reduction to an increase of
the buffer layer: ``{\it The conclusion is that, granted the law
of
 the wall and the defect law, a drag reduction must appear as a thickened
 sublayer...The sensitivity is such that doubling the sublayer thickness about
halves the skin friction}" (\cite{Lumley73}, p. 376). According to
Lumley, the increase of the buffer layer is due to an overall
increase in the fluid viscosity because of polymer stretching and
there should exist a critical polymer relaxation time $\tau$ for
drag reduction to occur. He also argued, without any quantitative
derivations, that the increase should be proportional to polymer
concentration for small concentration. At variance with Lumley's
interpretation, de Gennes suggested that polymers changes the
Kolmogorov cascade of turbulence at a critical scale where most of
the turbulent energy goes to polymer elastic stretching. The de
Gennes scale depends on polymer concentration and there should
exist a critical concentration for drag reduction to occur.
Nevertheless de Gennes argued: ``But the net result is still an
enhancement of the intermediate `buffer layer'.  We expect drag
reduction from this, although we have not carried out detailed
analog Lumley' matching" (\cite{deGennes}, p. 35). It is
worthwhile to mention that de Gennes argued that the critical
scale for polymer stretching should act as effective cutoff scale
for turbulent fluctuations, i.e., an effective Kolmogorov scale,
although de Gennes stated ``Thus we do not know the ultimate fate
of the turbulent energy" (\cite{deGennes}, p. 44). Finally, de
Gennes never mentioned drag reduction
 as induced by an energy flux that went from small scale polymer fluctuations to turbulent fluctuations.
 At any rate, neither de Gennes nor Lumley was able to provide the explanation
 of why the buffer layer should increase due to polymer stretching. Also, at the time
 when Lumley and de Gennes theories were proposed, no DNSs based on
 the FENE-P or Oldroyd-B models were available.
Our task is now to provide a quantitative explanation for drag
reduction and MDR within the framework of the FENE-P or Oldroyd-B
models, in which the ultimate fate of turbulent
 kinetic energy, i.e., the energy dissipation, is given by Equation \ref{4.4}.
 One possible answer to our question is provided in References \cite{Procaccia2008} and
 \cite{Benzi2006},
 and it is based on the idea that the polymer stretching produces an effective space-dependent viscosity.
 The phenomenological theory is based on the following observations:
\begin{enumerate}
\item {In wall-bounded turbulent flows, where the local shear $S$
is large,  the polymer conformational tensor satisfies the
inequalities $R_{xx} \gg R_{xy} \gg R_{yy} \sim R_{zz}$.} \item
{To first order approximation one can estimate $R_{xx} \sim S
\tau_{\rm eff} R_{xy}$ and $R_{xy} \sim S \tau_{\rm eff} R_{yy}$
where $\tau_{\rm eff} $ is given in Equation \ref{tau_eff}.} \item
{Turbulent kinetic energy $K(y)$ is fixed by the Lumley-de Gennes
criteria $\sqrt{K} \sim y/\tau_{\rm eff}$.} \item{ Turbulent
energy dissipation is the sum of the viscous effect and
$\nu_p(R_{yy}+R_{zz})/\tau_{\rm eff}^2 \sim \nu_p K(y) R_{yy} /
y^2 $. }
\end{enumerate}
Point 2 implies that that  terms $ \langle R_{xk} \partial_k u'_x
\rangle$ can be neglected, where $u'_x$ is the turbulent velocity
fluctuation in $x$-direction. Based upon  the previous points, one
can show that Equations \ref{5.20} and \ref{5.21} are modified as
\begin{eqnarray}
\label{5.23}
\nu^+(y^+) S^+ + W^+ = 1 \\
\label{5.24} \nu^+(y^+) \left( \frac{\Delta^+}{y^+} \right)^2 +
\frac{\sqrt{W^+}}{\kappa_v y^+} = S^+
\end{eqnarray}
where $ \nu^+ = 1 +  A R_{yy}$, and $A$ is a constant depending on
the ratio $\nu_p / \nu_s$ and $f(R)$, and $\Delta^+$ is a constant
that approaches $\delta^+$ in the limit concentration $c \to 0$.
We remark that the very existence of a space-dependent viscosity
can be numerically checked by computing the quantity $g(y) =
\epsilon_T/ \nu_0 \langle (\nabla {\bf u})^2 \rangle$, where
$\epsilon_T$ is the total  energy dissipation in the system and
$\nu_0 \langle (\nabla {\bf u})^2 \rangle$ is the energy
dissipation due to velocity fluctuations. It turns out that
$g(y)-1$ is a linear function of $y$ with very good accuracy (see,
for instance Figure \ref{fig5}b). Also, the existence of a
space-dependent viscosity can be derived in the generalization of
the classical Prandtl-Blasius boundary layer theory for the
Oldroyd-B or FENE-P models, as recently shown in References
\cite{Blasiusjfm,FENEPjfm}~(see Section 6). The approach described
in Reference \cite{Procaccia2008} justifies two independent
questions: (a) Does space-dependent viscosity produce drag
reduction? (b) What is  the functional form of $\nu^+(y^+)$ that
maximizes drag reduction and, eventually, provide a quantitative
prediction of the MDR asymptote? In principle, both questions can
be answered using numerical simulations. In Reference
\cite{Deangelis2004b},  linear viscosity profiles are shown to
produce drag reduction in wall-bounded turbulent flows and, for
increasing slope of the space-dependent viscosity, one observes a
decrease in the drag. Concerning the second question, using
Equations \ref{5.23} and \ref{5.24}, Reference \cite{Benzi2005b}
showed that the MDR can be achieved when $\nu^+(y^+)$ is linear
and that the resulting velocity profile is given by
\begin{equation}
\label{5.25} V^+(y^+) = 12 \ln(y^+) - 17.8
\end{equation}
assuming $\delta^+ \sim 6 $. Equation  \ref{5.25} is plotted in
Figure \ref{fig2}b and it is extremely close to the MDR behavior
discussed in Reference \cite{Virk75}. It is interesting to remark
that the asymptotic behavior given by Equation \ref{5.25} is
achieved in the limit $W^+ \rightarrow 0$, i.e., in the limit
turbulence tends to vanish. In this limit, therefore, the Lyapunov
exponent of the system becomes extremely small and, consequently,
the correlation time of any fluctuating quantity becomes very
long. This phenomenon is observed in DNS of the FENE-P model as
discussed in Reference \cite{Graham2012}, where periods of
hibernation and activation of turbulence activity are observed
near MDR. We emphasize that the formulation of the MDR as a
variational problem for the space-dependent viscosity implies the
universality of the MDR behavior, independent of polymer
characteristics. For very large Wi there may be non-negligible
contributions from correlations between the fluctuating strain
rates $\partial_i u'_j$ and the conformational tensor $R_{ij}$.
These contributions may support turbulent fluctuations near the
MDR asymptote in a way similar to the effect of elastic
turbulence, i.e., there may be on average a flux of energy from
polymer to turbulence. This possibility was pointed out in
References \cite{Dubief2012} and \cite{Dubief2013} and referred to
as elasto-inertial turbulence. However, this effect may be
consistent with the existence of a space-dependent viscosity
described in Equations  {\ref{5.23} and \ref{5.24} . Also, for
homogeneous shear flow simulations reported in \cite{Brasseur2010}
as well as in the numerical simulations of wall-bounded turbulence
discussed in Reference \cite{Ptasinski2003}, no energy flux from
polymer to turbulence has been found.

It is worth noting that asymptotic behavior similar to MDR is
observed in  numerical simulations of the FENE-P model in
homogeneous shear turbulence \cite{Brasseur2010}, where the
concept of a space-dependent viscosity cannot work. However, the
concept of space-dependent viscosity can be generalized to
scale-dependent viscosity as discussed in Reference
\cite{Benzi2004c} and a scale-dependent viscosity may explain drag
reduction for the cases reviewed in Section 5.1.

\section{POLYMER IN HEAT TRANSFER FLOWS}

The effect of polymers on heat transfer has been studied in
turbulent Rayleigh-B\'{e}nard (RB) convection. In the RB system, a
fluid is constrained between two horizontal plates that are heated
from below and cooled from above, and the system is controlled by
two parameters: the Rayleigh number, Ra $ = \alpha g \delta T
H^3/(\kappa \nu_s)$, which measures the thermal forcing due to the
temperature difference $\delta T$ between the two plates, and the
Prandtl number, Pr $ =\nu_s/\kappa$, which is the ratio between
the kinematic viscosity $\nu_s$ and the thermal diffusivity
$\kappa$ of the fluid. In addition, $\alpha$ is the isobaric
volume expansion coefficient of the fluid, $g$ the acceleration
due to gravity, and $H$ is the vertical distance between the top
and bottom plates. In turbulent RB convection, there are distinct
flow regions, namely viscous boundary layers near all rigid walls
and two thermal boundary layers, one above the bottom plate and
one below the top plate, and an approximately homogeneous bulk
flow in the central region of the convection cell. In the
Boussinesq approximation, the equations of motion for RB
convection with polymers are
\begin{eqnarray}
\partial_t u_i +  u_j \partial_j u_i &=& - \frac{1}{\rho} \partial_i p +  \partial_j T_{ij} + \nu_s \Delta u_i +
 \alpha g (T-T_*)\ \delta_{iz}
\label{6.1} \\
\partial_t T +  u_j \partial_j T &=&  \kappa \Delta T \label{6.2}
\end{eqnarray}
where $T$ is the temperature field, $T_*$ is the mean temperature
averaged over time and the whole system, $z$ is along the vertical
direction, and the polymer stress tensor $T_{ij}$ is given by
Equation \ref{4.2} with the conformation tensor $R_{ij}$ governed
by Equation \ref{1.2}. Heat transport is measured by the Nusselt
number (Nu), which is the normalized heat flux defined by
\begin{equation}
{\rm Nu} \equiv \frac{\langle u_z T - \kappa \partial_z T
\rangle_A}{\kappa \delta T/H} \label{Nu}
\end{equation}
where $\langle \cdots \rangle_A$ denotes an average over a
horizontal plane of the convection cell and time.

\begin{figure}[h]
\includegraphics[width=4in]{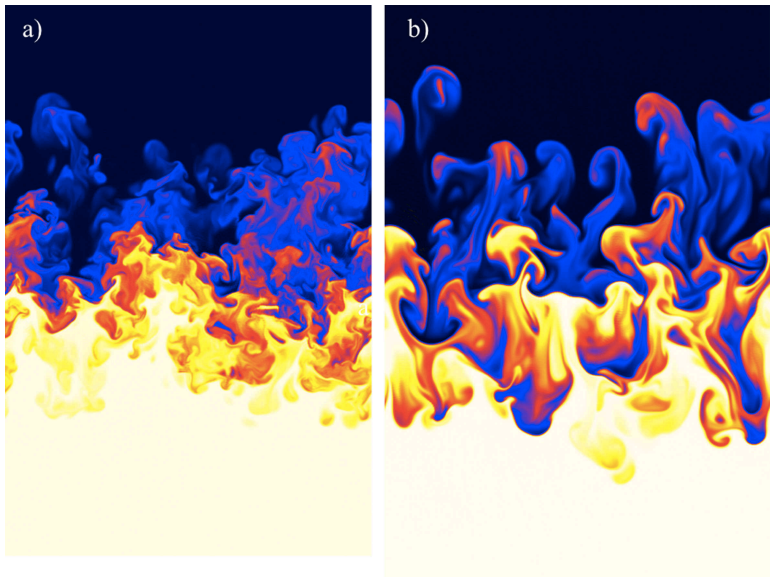}
\caption{Snapshots of the temperature field  (a) without and (b)
with polymers in direct numerical simulation of Rayleigh-Taylor
turbulence. Due to polymers, the turbulent plumes become larger in
size and small scales fluctuations are suppressed. As a result,
the heat flux increases. Reproduced from
Reference~\cite{Boffetta2010} with permission.} \label{fig6}
\end{figure}

The bulk flow of turbulent RB convection without the boundary
layers is believed~\cite{LohseToschiPRL} to be a good
approximation of the ultimate regime at large Ra. The effect of
polymer in this regime has been studied by DNS of the equations of
motion with periodic boundary conditions~\cite{BCEPRL}.  In this
case, polymer enhances the length scale $l_T$ of thermal plumes
and heat transfer increases as Nu~$\sim l_T^{3/2} \sim {\rm
Wi}^{3/2}$, where Wi $ \equiv \tau
 U_c/H$ and $U_c = \sqrt{\alpha g \delta T H}$. A similar enhancement in heat
transfer by polymers has been found in Rayleigh-Taylor
turbulence~\cite{Boffetta2010,Boffetta2017} (see Figure
\ref{fig6}). In contrast, an experimental
investigation~\cite{Ahlers} of turbulent RB convection at moderate
Ra reported a small but clear reduction of Nu in the presence of
polymers, and the amount of heat reduction increases with polymer
concentration. At low Ra, most of the energy dissipation is
concentrated in the boundary layers, and thus the numerical
results for the ultimate regime do not apply. In later
experimental studies~\cite{WeiPRE,XiaJFM}, both a convection cell
with usual smooth top and bottom plates and a convection cell with
rough top and bottom plates have been used. The reduction of heat
transport by polymers at moderate Ra has been confirmed in the
smooth cell, whereas an enhancement in heat transfer is observed
in the rough cell when the polymer concentration is not too
small~(see Figure \ref{fig7}). It is believed that the pyramidal
structures of the rough plates perturb the boundary layers and
make the flow resembling that of the bulk flow even at moderate
Ra.

\begin{figure}[h]
\includegraphics[width=5in]{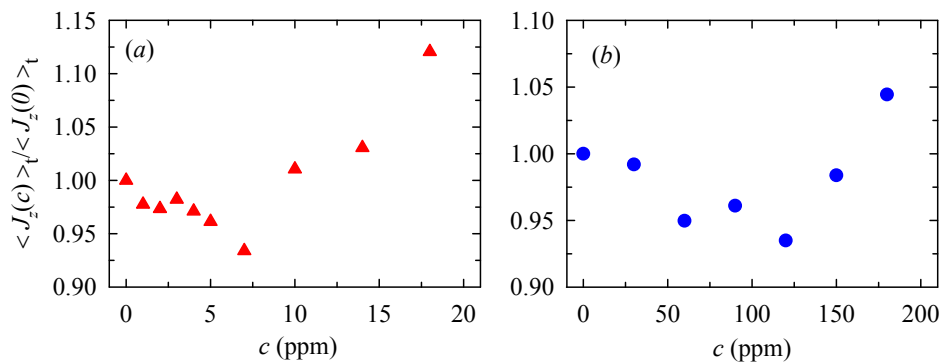}
\caption{Time-averaged local vertical heat flux $\langle J_z(c)
\rangle_t \equiv \langle u_z(t)[T(t)-T_*] \rangle_t H/(\kappa
\delta T)$ at the center of the rough convection cell, normalized
by its Newtonian value $\langle J_z(0) \rangle_t$ measured at the
same Ra, as a function of polymer concentration $c$ for (a)
polyacrylamide and (b) polyethylene oxide. Adapted from
Reference~\cite{XiaJFM} with permission.} \label{fig7}
\end{figure}

In an attempt to understand the experimental results at moderate
Ra in the usual smooth convection cell, the effect of polymers in
the framework of the classical Prandtl-Blasius boundary layer
theory was analyzed~\cite{Blasiusjfm,FENEPjfm}, which is
known~\cite{GL2000,Revmodlohse} to be a good approximation for
moderate Ra and stable boundary layers. The outcomes of the
approach discussed in References \cite{Blasiusjfm,FENEPjfm} are
the following: First, the effect of polymers induces an effective
space-dependent viscosity in the equation of motion, which peaked
in the boundary layer and vanishes away from the boundary layer.
Second, upon increasing the polymer concentration, heat reduction
is observed, i.e., Nu decreases with respect to ${\rm Nu}_0$
corresponding to the Newtonian flow with no polymer at the same
zero-shear viscosity. Finally, it can be shown that if the
effective viscosity extends into the bulk of the system, i.e., the
center of the cell, the value of Nu increases against ${\rm
Nu}_0$. Physically this means that heat enhancement should be
observed if a substantial stretching of the polymer in the bulk of
the system exists. The last statement has been confirmed by a very
recent DNS of turbulent RB convection  with polymers~\cite{BCEPRE}
at moderate Ra and $2 \le {\rm Wi} \le 50$.

\section{CONCLUSIONS}

Polymer-flow interactions lead to a number of non-trivial and
counterintuitive phenomena observed in laboratory experiments,
namely drag reduction, elastic turbulence and modification of heat
transport in natural convection. Our present knowledge is mostly
based on direct numerical simulations, which provide a qualitative
(and in some cases quantitative) explanation of experimental
results. Still many different questions should be answered in
order to define a possible unified scenario (if it exists). In
this review, we have highlighted  the most relevant results
obtained in the last twenty years trying to capture the basic
features underlying different theoretical approaches and
explanations. It is fair to say that much progress have been made
by suitable analyses of both laboratory and numerical
investigations. Such progress has been able to foster new
theoretical interpretations and to open a new view on the subject.
The same approach can be applied to study the interaction of
turbulent and/or laminar flows for other complex fluids as in the
case of rigid polymers and surfactants. The outcomes of such
investigations will be relevant for both our basic scientific
understanding of complex fluid dynamics and technological
applications.

\section*{SUMMARY POINTS}
\begin{enumerate}
\item[1.] FENE-P and Oldroyd-B models can explain qualitatively
most of the experimental results on drag reduction and elastic
turbulence. \item[2.] A clear cut definition of coil-stretch
transition has been successfully provided in terms of Lagrangian
properties of polymer dynamics. \item[3.] The notion of drag
reduction can be generalized to flows without boundaries.
\item[4.] Polymer-flow interaction can change the heat transport
in natural and forced convection.
\end{enumerate}

\section*{FUTURE ISSUES}
\begin{enumerate}
\item[1.]  More realistic models of flexible polymers are needed
to make quantitative predictions of polymer-flow interactions.
\item[2.]  Inertial effects on polymer dynamics, due to density
mismatch between polymer and flow, need to be investigated.
\end{enumerate}

\section*{DISCLOSURE STATEMENT}
The authors are not aware of any affiliations, memberships,
funding, or financial holdings that might be perceived as
affecting the objectivity of this review.

\section*{ACKNOWLEDGMENTS}
R.B. acknowledges funding from the European Research Council under
the European Union's Seventh Framework Programme, AdG ERC Grant
Agreement No. 339032 and E.S.C.C. acknowledges support from the
Hong Kong Research Grants Council (Grants no 400304, 400311).

\end{document}